# Synthesis and characterization of novel benzimidazole embedded 1,3,5-trisubstituted pyrazolines as antimicrobial agents

GOPAL K. PADHY[1,2], JAGADEESH PANDA[3] and AJAYA K. BEHERA[1]*

[1]*Organic Synthesis Laboratory, School of Chemistry, Sambalpur University, Jyoti Vihar, Burla 768019, India,* [2]*Maharajah's College of Pharmacy, Phool Baugh, Vizianagaram 535002, India and* [3]*Raghu College of Pharmacy, Dakamarri, Visakhapatnam 531162, India*

*Abstract*: Efficient syntheses of some new substituted pyrazoline derivatives linked to substituted benzimidazole scaffold were performed by multistep reaction sequences. All the synthesized compounds were characterized using elemental analysis and spectral studies (IR, 1D/2D NMR techniques and mass spectrometry). The synthesized compounds were screened for their antimicrobial activity against selected Gram-positive and Gram-negative bacteria, and fungi strain. The compounds with halo substituted phenyl group at C5 of the 1-phenyl pyrazoline ring (**15**, **16** and **17**) showed significant antibacterial activity. Among the screened compounds, **17** showed most potent inhibitory activity ($MIC$ = 64 μg mL$^{-1}$) against a bacterial strain. The tested compounds were found to be almost inactive against the fungal strain *C. albicans*, apart from pyrazoline-1-carbothiomide **21**, which was moderately active.

INTRODUCTION

In the current scenario, bacterial infections bring about a serious threat to human lives due to their rapid resistance to existing antibiotics. Thus, exploration of new types of antibacterial agents has become extremely vital. Benzimidazole derivatives are very useful for the development of molecules of pharmaceutical interest due to their pharmacological activities, including antimicrobial,[1–3] anticancer,[4–6] antidiabetic[7] and plasmin inhibitor.[8] Moreover, *N*-benzyl-substituted benzimidazoles have been synthesized showing promising antibacterial activities, among which amidine[9] **I** and 2-(piperidin-4-yl)benzimidazole[10] **II** are examples (Fig. 1).

---

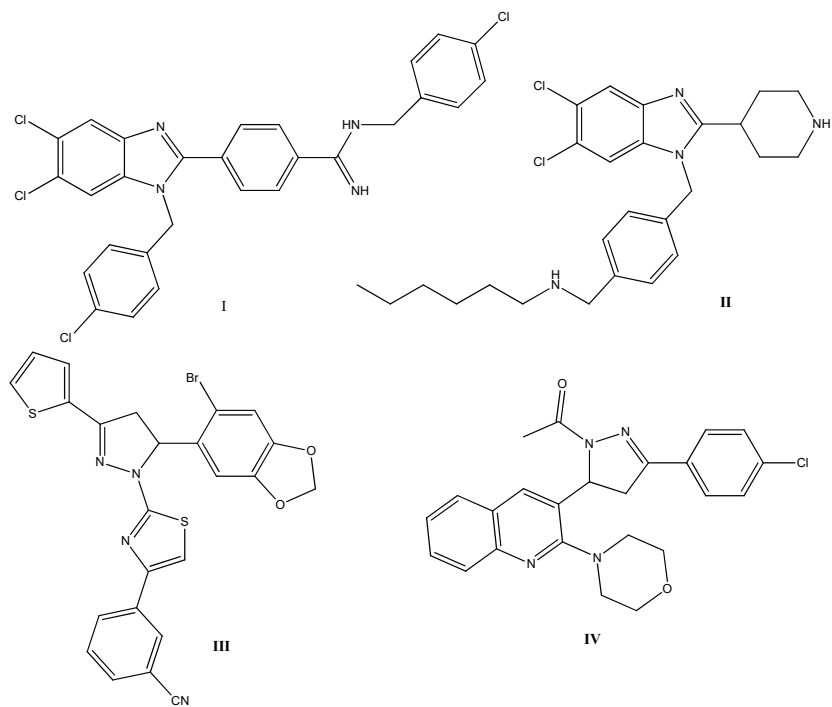

Fig. 1. Structures of some benzimidazoles and pyrazolines with antibacterial activity.

The pyrazoline motif is a core structure in numerous biologically active compounds. Some representatives of this heterocycle exhibited antimicrobial,[11,12] anticancer,[13,14] anti-inflammatory,[15,16] and monoamine oxidase inhibitory[17,18] activities. Series of novel thiazolyl-pyrazoline,[19] such as **III** and morpholinoquinoline clubbed pyrazoline[20] **IV** were recently reported as potent antibacterial agents (Fig. 1). The combination of *N*-benzylbenzimidazole and pyrazoline fragments in one molecule is expected to be a perspective approach to design promising antimicrobial agents. Thus, a novel series of *N*-benzyl attached benzimidazolyl pyrazolines was designed, synthesized and evaluated against different bacteria and fungi.

EXPERIMENTAL

*Chemistry*

The chemicals used were laboratory grade and procured from Merck (India), Fischer Scientific (India) and Finar (India). IR spectra were obtained on a Bruker ALPHA-T FT-IR spectrometer (KBr pellets using opus software). $^1$H- and $^{13}$C-NMR spectra were recorded on a Bruker AVANCE III 500 MHz (AV 500) spectrometer using TMS as an internal standard in DMSO-$d_6$/CDCl$_3$. The mass spectra were recorded on a Varian Inc 410 Prostar Binary LC––MS and an Agilent 6410 LC–MS spectrometer. Melting points were determined by the open tube capillary method and are uncorrected. Progress of the reaction and purity of the products was checked by thin layer chromatography (TLC). The spots were located under iodine vapours/UV light. The physical, analytical and spectral data for the compounds are given in the Supplementary material to this paper.

*General procedure for synthesis of 3-aryl-1-(1-benzyl-1H-benzo[d]imidazol-2-yl)-2-propen-1--ones (8–12)*

Chalcones **3**–**7** (10 mmol) were dissolved in 40 mL dry acetone and then anhydrous $K_2CO_3$ (15 mmol) was added to the solution. Later benzyl chloride (40 mmol) was added to the mixture and the contents were heated under reflux. The progress of the reaction was monitored by TLC (benzene–ethyl acetate, 4:1). After completion of the reaction (22–26 h), the reaction mixture was cooled and then poured into crushed ice. The obtained solid were filtered and recrystallized form alcohol. It should be noted that compounds **9**– **11** crystallised out from the reaction mixture after cooling, and they were then collected by filtration. Compounds **8** and **10** were previously reported.[21] However, neither of them has been examined for their antimicrobial activities.

*General procedure for the synthesis of 1-benzyl-2-(5-aryl-1-phenyl-4,5-dihydro-1H-pyrazol--3-yl)-1H-benzimidazoles (13–17)*

To a solution of chalcones **8**–**12** (2 mmol) in acetic acid, phenylhydrazine (3 mmol) was added dropwise. The reaction mixture was heated under reflux. The progress of the reaction was monitored by TLC (benzene–ethyl acetate, 5:1). After completion of reaction (6–8 h), the reaction mixture was cooled and then poured into crushed ice. The obtained solid pyrazolines were washed with diethyl ether and recrystallized from acetone.

*General procedure for synthesis of 5-aryl-3-(1-benzyl-1H-benzimidazol-2-yl)-4,5-dihydro--1H-pyrazole-1-carbothioamides (18–22)*

A mixture of chalcones **8**–**12** (2 mmol), thiosemicarbazide (6 mmol) and NaOH (4 mmol) was refluxed in ethanol (15 ml). The progress of the reaction was monitored by TLC (benzene–ethyl acetate, 5:1). After completion of reaction (4–6 h), the reaction mixture was cooled. The precipitate formed was filtered and washed with acetone.

*Microbiology*

The compounds **8**–**22** were evaluated for their *in vitro* antimicrobial activity against bacterial strains, *viz. Bacillus subtilis* MTCC 441, *Staphylococcus aureus* MTCC 3160, *Pseudomonas aeruginosa* MTCC 4673 and *Escherichia coli* MTCC 739, and the fungi *Candida albicans* MTCC 183. The reference cultures were procured from the Institute of Microbial Technology (IMTECH), Chandigarh, India-160036. Minimum inhibitory concentrations (*MIC*) were determined using nutrient broth (NB) for the bacteria and Sabouraud dextrose broth (SDB) for fungi by the two-fold serial dilution method.[22,23] The cultures were incubated for 24 h for the bacteria and 48 h for the fungi at 35 °C and the growth was monitored. The lowest concentration required to arrest the growth of microorganism was regarded as the minimum inhibitory concentration (*MIC*). Ciprofloxacin and fluconazole were used as positive control for the bacteria and fungi, respectively. DMSO was used as the negative control. The determinations of the antimicrobial activities of the compounds were performed in duplicate.

RESULTS AND DISCUSSION

*Chemistry*

The target compounds described in this study were prepared as outlined in Scheme 1. Condensation of *o*-phenylenediamine with lactic acid under Phillips conditions led to 2-(1-hydroxyethyl)benzimidazole (**1**), chromic oxidation of the latter followed by neutralization with ammonia led to 2-acetylbenzimidazole (**2**).[24] The required synthons 3-aryl-1-(1*H*-benzimidazol-2-yl)-2-propen-1-ones (**3**–**7**) were prepared by Claisen–Schmidt condensation of 2-acetylbezimidazole with substituted aromatic aldehydes in presence of NaOH.[25] Condensation of the 3-aryl-1-benzimidazolyl-2-propen-1-one derivatives **3**–**7** with benzyl chloride gave the corresponding 3-aryl-1-(1-benzyl-1*H*-benzimidazol-2-yl)-2-propen-1--ones **8**–**12**. The reaction of **8**–**12** with phenylhydrazine in the presence of acetic acid afforded 1-benzyl-2-(5-aryl-1-phenyl-4,5-dihydro-1*H*-pyrazol-3-yl)-1*H*-benzimidazoles **13**–**17**, whereas when **8**–**12** were condensed with thiosemicarbazide in presence of NaOH, 5-aryl-3-(1-benzyl-1*H*-benzimidazol-2-yl)-4,5-dihydro--1*H*-pyrazole-1-carbothioamides **18**–**22** were obtained in good yields.

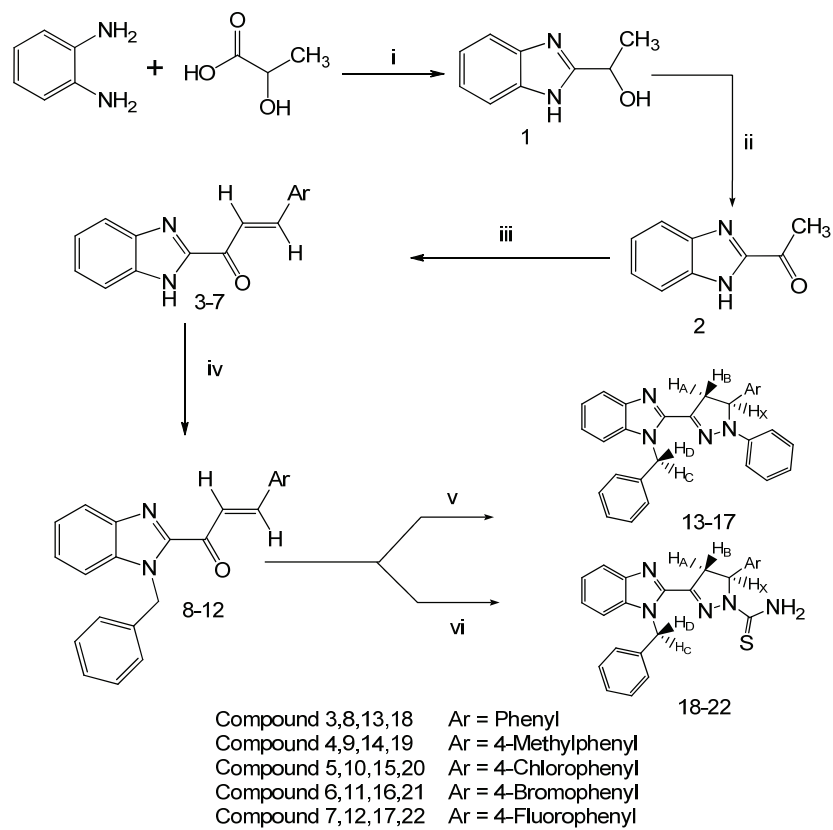

Scheme 1. Synthetic route to 1-benzyl-2-(1-substituted-5-aryl-4,5-dihydro-1*H*-pyrazol-3-yl)--1*H*-benzimidazoles. Reagents and conditions: i) 4 M HCl, reflux, 8 h; ii) $K_2Cr_2O_7$, dil. $H_2SO_4$, r.t., 2h; iii) Ar-CHO, 10 % aq NaOH, ethanol, r.t., 4–8 h; iv) benzyl chloride, dry acetone, anhydrous $K_2CO_3$, reflux, 22–26 h; v) phenylhydrazine, acetic acid, reflux, 6–8 h; vi) thiosemicarbazide, NaOH, ethanol, reflux, 4–6 h.

The structure of the synthesized compounds **8–22** were assigned based on elemental and spectroscopic analysis, IR, $^1$H-NMR, $^{13}$C-NMR and mass spectrometry (Supplementary material to this paper). The IR data were very informative and fully supported the proposed structures of the reported compounds. In the IR spectra of chalcones **8–12**, the (C=O) stretching was found in the expected region at 1652–1662 cm$^{-1}$. Furthermore, the presence of (C=N) and (C–N) stretching frequencies at 1594–1597 cm$^{-1}$ and 1283–1330 cm$^{-1}$ in the IR spectra of **13–22** confirmed the subsequent cyclization of the chalcone to the pyrazoline derivatives. In addition, derivatives **18—22** showed typical absorption bands due to (–NH$_2$) at 3290–3355 cm$^{-1}$ and 3218–3250 cm$^{-1}$.

The $^1$H-NMR spectrum compounds **8–12** exhibited two doublets with *J* values between 15–20 Hz, confirming the *trans* coupling and indicating the olefinic protons in the *E* form. The $^{13}$C-NMR spectra of **8–12** confirmed the presence of the *α,β*-unsaturated carbonyl system of chalcones by the presence of a peak at δ 182.40–182.53 ppm, corresponding to the propenone C1.

Although compounds **13–22** were racemates, their structures were unambiguously assigned with the help of spectral studies. The $^1$H-NMR spectra of compounds **13–22** displayed three sets of signals with an ABX pattern for the pyrazoline ring protons. The sterochemical nature of the hydrogens H$_A$, H$_B$ and H$_X$ was ascertained from a study of the coupling constants (*J*). The vicinal coupling constant between H$_A$ and H$_X$ was found to be 2.7–10.0 Hz ($J_{AX}$), which indicates that these hydrogens are *cis* to each other, while the *trans* relationship between H$_B$ and H$_X$ was evident from the coupling constants of $J_{BX}$ in the 10.0––15.0 Hz range. The coupling value of $J_{AB}$ in 17.5–20.0 Hz range between H$_A$ and H$_B$ evidently indicates their geminal placement at C4. The CH$_2$ protons of pyrazoline are diastereotopic and appeared as a pair of doublets of doublets at δ 2.90–3.10 ppm (H$_A$) and δ 3.85–4.05 ppm (H$_B$), due to the vicinal coupling with the H$_X$ proton and the geminal coupling with each other. The CH proton (H$_X$) appeared as a doublet of doublets at δ 5.72–5.92 ppm due to vicinal coupling with the two magnetically non-equivalent protons of the methylene group (H$_A$ and H$_B$) at position 4 of the pyrazoline ring. The two methylene protons of benzyl group on geminal coupling showed two separate doublets at δ 5.33–6.08 ppm (H$_C$) and δ 5.37–6.17 ppm (H$_D$), indicating the diastereotopic nature of the methylene hydrogen. This was further confirmed by the appearance of three sets of signals at δ 40.03–44.45 ppm (C4 of pyrazoline), δ 47.01–48.18 ppm (CH$_2$ of benzyl) and δ 57.54–62.62 ppm (C5 of pyrazoline) in the $^{13}$C-NMR spectra of pyrazolines. It should be noted that in the spectra of compounds **15** and **17,** the signal of C4 of pyrazoline was overshadowed by residual peaks of DMSO-*d$_6$*. Compounds **18–22** exhibited two separate broad singlets at δ 7.81–7.89 ppm and δ 8.26–8.38 ppm for the –NH$_2$ protons. All the other additional peaks observed were in agreement with the respective aromatic substituents and benzimidazole ring. The mass spectra and elemental analyses were also in agreement with the proposed structures.

Furthermore, the carbon–proton correlation of pyrazoline **13** was confirmed by an HSQC experiment. The $H_A$ and $H_B$ protons ($\delta$ 3.06 and 3.71 ppm) showed a cross-peak with the carbon signal at 40.03 ppm, which confirmed that this signal is due to the C4 carbon. Similarly, the doublet of doublets at $\delta$ 5.75 ppm is strongly correlated with the carbon resonance at $\delta$ 60.36 ppm. From this, it is inferred that the carbon signal is due to the C5 carbon. The carbon signal at $\delta$ 47.35 ppm correlated well with the two methylene protons ($\delta$ 5.34 and 5.39 ppm) of the benzyl group, which established the diastereotopic nature of the methylene hydrogens.

*Antimicrobial activity*

The synthesized compounds **8**–**22** were screened for their *in vitro* antimicrobial activity. The results of antimicrobial activities of the benzimidazole derivatives are presented in Table I. The antimicrobial screening data revealed that all the newly synthesized compounds exhibited weaker antimicrobial activities compared to those of the control drugs. For bacterial strains, the *MIC* values of the compounds ranged between 128–1024 µg mL$^{-1}$ for the chalcones **8**–**12**, between 64–1024 µg mL$^{-1}$ for the 1-phenylpyrazolines **13**–**17** and between 128––512 µg mL$^{-1}$ for the pyrazoline-1-carbothioamides **18**–**22**. Among all the tested compounds, compound **17** showed good activity (64 µg mL$^{-1}$) against the tested bacterial strains. The *MIC* value of ciprofloxacin was 6.25–12.5 µg mL$^{-1}$ for all the tested bacterial strains. Chalcones **8**–**12** and 1-phenylpyrazolines **13** and **14** were almost inactive against *C. albicans* but pyrazoline-1-carbothioamide **21** showed moderate activity against the fungi strain. The *MIC* value of fluconazole was 6.25 µg mL$^{-1}$ for the fungi *C. albicans*.

TABLE I. *In vitro* antimicrobial activity of the synthesized compounds **8**–**22**, *MIC* / µg mL$^{-1}$

| Compound | Gram-positive bacteria | | Gram-negative bacteria | | Fungi |
|---|---|---|---|---|---|
| | *S. aureus* | *B. subtilis* | *E. coli* | *P. aeruginosa* | *C. albicans* |
| **8** | 512 | 512 | >1024 | >1024 | >1024 |
| **9** | >1024 | 512 | >1024 | >1024 | >1024 |
| **10** | 256 | 256 | 512 | 512 | >1024 |
| **11** | 128 | 256 | 256 | 512 | >1024 |
| **12** | 128 | 128 | 256 | 256 | >1024 |
| **13** | 256 | 512 | 512 | 256 | >1024 |
| **14** | >1024 | 512 | >1024 | >1024 | >1024 |
| **15** | 64 | 128 | 256 | 128 | 512 |
| **16** | 128 | 64 | 256 | 128 | 512 |
| **17** | 64 | 128 | 64 | 256 | 256 |
| **18** | >1024 | 512 | >1024 | >1024 | 512 |
| **19** | >1024 | >1024 | 512 | >1024 | 512 |
| **20** | 512 | 256 | >1024 | 512 | 512 |
| **21** | 256 | 128 | 512 | 512 | 128 |
| **22** | 128 | 128 | 256 | 512 | 256 |
| Ciprofloxacin | 6.25 | 6.25 | 6.25 | 12.5 | – |
| Fluconazole | – | – | – | – | 6.25 |

Subsequently, preliminary SAR studies were performed to deduce how the structure variation and modification could affect the antimicrobial activity. The antimicrobial activities of the compounds are related to the presence of electron withdrawing or donating substituents on the benzene ring. Compounds containing electron withdrawing –F, –Cl and –Br exhibited good antimicrobial activity, whereas non-substituted compounds and compounds substituted with electron donating groups did not exhibit the same potency. For example, compound **17** showed an *MIC* of 64 µg mL$^{-1}$ against *S. aureus* and *E. coli*, whereas compound **13** showed an *MIC* of 256 and 512 µg mL$^{-1}$, respectively. The reason for above antibacterial activities could be explained by electron density, which plays an important role for the optimum activity.

## CONCLUSIONS

In this paper, the synthesis and characterization of three new chalcones, five 1-phenylpyrazolines and five pyrazoline-1-carbothioamides containing the *N*-benzylbenzimidazole moiety are presented. The structures of the new compounds were confirmed by spectral data (IR, $^1$H-NMR, $^{13}$C-NMR and mass spectrometry) and elemental analysis. All the compounds were investigated for their antimicrobial activity against *B. subtilis*, *S. aureus*, *P. aeruginosa*, *E. coli* and *C. albicans*. The data indicated weak antibacterial activity, except for compound **17** (which presented good activity against *S. aureus* and *E. coli*), **16** (which presented moderate activity on *S. aureus* and *P. aeruginosa* and **15** (which presented moderate action on *B. subtilis* and *P. aeruginosa*). The weak antibacterial activity could be because the tested compounds were in a racemic form. Based on the *MIC* values presented by the tested compounds, it could be concluded that, in general, the derivatives containing a fluorine, chlorine and bromine atom had better antibacterial activity against the tested strains. The tested compounds were found to be either inactive, or moderately active (**21**), against the fungal strain *C. albicans*.

## SUPPLEMENTARY MATERIAL

Analytical and spectral data of the synthesized compounds are available electronically at the pages of the journal website: http://www.shd.org.rs/JSCS/, or from the corresponding author on request.

*Acknowledgements*. The authors gratefully acknowledge the Sophisticated Instrumentation Facility (SAIF), IIT Madras and SAIF, IIT Bombay, India, for providing the spectral data. Gopal is thankful to Sri. P. Ashok Gajapathi Raju, Chairman, MANSAS, Vizianagaram and Dr. P. Udaya Shankar, Principal Maharajah's College of pharmacy, Vizianagaram for providing the necessary infrastructure and facility.

ИЗВОД
СИНТЕЗА И КАРАКТЕРИЗАЦИЈА НОВИХ БЕНЗИМИДАЗОЛА КОЈИ САДРЖЕ 1,3,5-ТРИСУПСТИТУИСАНИ ПИРАЗОЛИН КАО АНТИМИКРОБНИХ ЈЕДИЊЕЊА

GOPAL K. PADHY[1,2], JAGADEESH PANDA[3] и AJAYA K BEHERA[1]

[1]*Organic Synthesis Laboratory, School of Chemistry, Sambalpur University, Jyoti Vihar, Burla 768019, India,* [2]*Maharajah's College of Pharmacy, Phool Baugh, Vizianagaram 535002, India* и [3]*Raghu College of Pharmacy, Dakamarri, Visakhapatnam 531162, India*

Извршена је ефикасна синтеза нових супституисаних деривата пиразолина везаних са бензимидазолским прстеном, применом вишефазне реакционе секвенције. Сва синтетисана једињења окарактерисана су елементалном анализом, спектроскопским методама (IR, 1D/2D NMR) и масеном спектрометријом. Испитана је антимикробна активност синтетисаних једињења према одабраним сојевима Грам-позитивних и Грам-негативних бактерија и одабраних сојева гљивица. Једињења која поседују халоген-супституисану ароматичну групу на C5 1-фенилпиразолинског прстена (**15**–**17**) показују значајну антибактеријску активност. Од испитиваних једињења, дериват **17** показује највећу инхибиторну активност ($MIC = 64$ μg mL$^{-1}$). Испитивана једињења су показала готово потпуно одсуство активности према *C. albicans*, осим пиразолин-1-карботиоамида **21** који је показао умерену активност.

(Примљено 4. јуна 2016, ревидирано 3. априла, прихваћено 17. јула 2017)